\documentclass[a4paper]{spie}

\usepackage[]{graphicx}
\usepackage{aas_macros}
\usepackage[]{subfigure}

\title{A versatile facility for the calibration of X-ray polarimeters with polarized 
and unpolarized controlled beams} 

\author{Fabio~Muleri\supit{a,b}, Paolo~Soffitta\supit{a}, Ronaldo~Bellazzini\supit{c},
Alessandro~Brez\supit{c}, Enrico~Costa\supit{a}, Massimo~Frutti\supit{a},
Marcello Mastropietro\supit{a,d}, Ennio Morelli\supit{a,e}, Michele~Pinchera\supit{c},
Alda~Rubini\supit{a}, Gloria~Spandre\supit{c}
\skiplinehalf
\supit{a} Istituto di Astrofisica Spaziale e Fisica Cosmica, Via del Fosso del Cavaliere 100,
I-00133 Roma, Italy;
\\
\supit{b} Universit\`{a} di Roma Tor Vergata, Dipartimento di Fisica, via della Ricerca Scientifica
1, 00133 Roma, Italy
\\
\supit{c} Istituto Nazionale di Fisica Nucleare, Largo B. Pontecorvo 3, I-56127 Pisa,  Italy
\\
\supit{d} CNR, Istituto Metodologie Inorganiche e dei Plasmi, Area Ricerca Montelibretti, Italy
\\
\supit{e} Istituto di Astrofisica Spaziale e Fisica Cosmica, Via Gobetti 101, I-40129 Bologna, Italy
}

\authorinfo{Further author information: (Send correspondence to Fabio Muleri)
\\Fabio Muleri: E-mail: Fabio.Muleri@iasf-roma.inaf.it,
Telephone: +39-0649934565}

\begin{document} 
\maketitle 

\begin{abstract}
We devised and built a versatile facility for the calibration of the next generation X-ray
polarimeters with unpolarized and polarized radiation. The former is produced at 5.9 keV by means of
a Fe$^{55}$ radioactive source or by X-ray tubes, while the latter is obtained by Bragg
diffraction at nearly 45~degrees. Crystals tuned with the emission lines of X-ray tubes with
molybdenum, rhodium, calcium and titanium anodes are employed for the efficient production of highly
polarized photons at 2.29, 2.69, 3.69 and 4.51~keV respectively. Moreover the continuum emission is
exploited for the production of polarized photons at 1.65 keV and 2.04~keV and at energies
corresponding to the higher  orders of diffraction. The photons are collimated by means of
interchangeable capillary plates and diaphragms, allowing a trade-off between collimation and high
fluxes. The direction of the beam is accurately arranged by means of  high precision motorized
stages, controlled via computer so that long and automatic measurements can be done. Selecting the
direction of polarization and the incidence point we can map the response of imaging devices to both
polarized and unpolarized radiation. Changing  the inclination of the beam we can study the
systematic effects due to the focusing of grazing incidence optics and the feasibility of
instruments with large field  of view.
\end{abstract}


\section{Introduction}

The continuous development of new polarimeters based on the photoelectric effects has recently
renewed the interest in the study of polarization from astrophysical sources in the X-ray band. The
Gas Pixel Detector, developed by the INFN of Pisa\cite{Costa2001, Bellazzini2006, Bellazzini2007},
is one of the most advanced project in this field. It allows a large increase of sensibility with
respect to the previous instruments, based on Bragg diffraction or Thomson scattering, and makes
feasible the development of the X-ray polarimetry of astrophysical sources, for the first time after
the pioneer experiments in the '70.

The current version of the GPD works between 2 and 10~keV, with a sensitivity peaking at about
3~keV. Its response has been studied by employing a Monte Carlo software, while its calibration
was at first attained with a polarized source based on Thomson scattering at nearly 90~degrees on a
lithium target, enclosed in beryllium to prevent its oxidation and nitridation. However, such a
Thomson scattering source doesn't allow the production of photons below about 5~keV, i.e. in the
range of maximum sensitivity of the GPD, because of the photoelectric absorption in the target.

The construction by IASF/INAF of Rome of a very compact X-ray polarized source, based on the Bragg
diffraction at nearly 45 degrees \cite{Muleri2007}, has eventually overcome this problem and
allowed for the first time the measurement of the sensibility of the GPD at low energy. We
employed mosaic graphite and flat aluminum crystals to produce nearly completely polarized photons
at 2.6, 3.7, and 5.2~keV from the diffraction of unpolarized continuum or line emission. These
measurements have been proved that the GPD responds as expected on the basis of a Monte Carlo
software \cite{Muleri2008}.

The next step in the characterization of the GPD is then the systematic study of its response with a
space-controlled polarized beam, with energies covering the whole response of the instruments. Here
we present the facility we devised and built at this aim. 

\section{The production of highly polarized photons}

\subsection{The Bragg Diffraction}

Low energy and polarized photons can be produced by means of Bragg diffraction at nearly 45~degrees.
If $P_E(\theta)$ is the diffracted intensity when parallel monochromatic radiation of unit
intensity and energy $E$ is incident at glancing angle $\theta$ to the diffracting planes of a flat
crystal, the integrated reflectivity $R_E$ is defined as \cite{Evans1977}:
\begin{equation}
R_\lambda = \int_0^\frac{\pi}{2} P_\lambda \left(\theta\right) d\theta.
\end{equation}

\begin{figure}[htbp]
\begin{center}
\includegraphics[angle=0, totalheight=6cm]{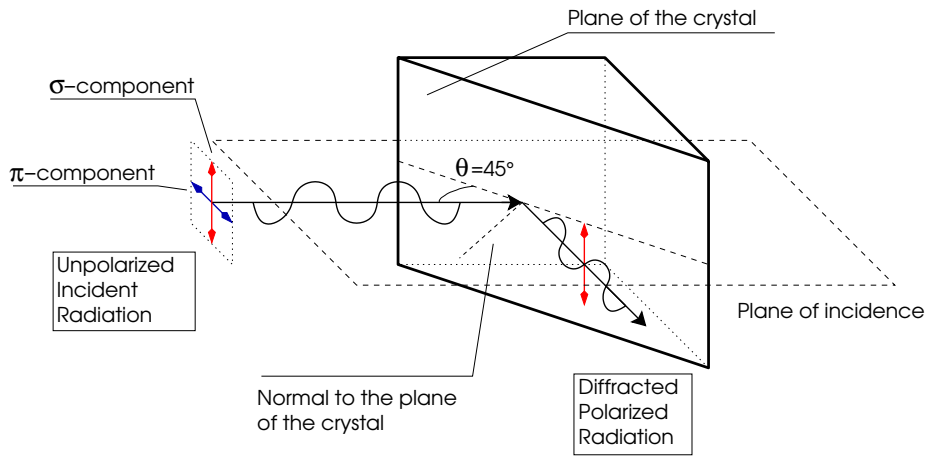}
\end{center}
\caption{\small Geometry of the Bragg diffraction at 45 degrees. Unpolarized radiation is polarized
because the component polarized parallel to the incidence plane, defined by the plane where the
direction of the incident radiation and the normal of diffraction planes lie, is more efficiently
absorbed with respect to component polarized orthogonally to the incidence plane.
\label{fig:BraggDiffraction}}
\end{figure}

The integrated reflectivity expresses the ``efficiency'' of the diffraction and is sensitive to the
polarization of the incident radiation. If we define the ratio $k={R_E^\pi}/{R_E^\sigma}$ of the
integrated reflectivity for radiation polarized parallel ($\pi$-component) and perpendicularly
($\sigma$-component) to the incidence plane, $k$ depends on the diffraction angle and is
generally lower than 1 (see Fig.~\ref{fig:BraggDiffraction}). The radiation perpendicularly
polarized to the incidence plane is then diffracted more efficiently and the net effect is that the
diffracted radiation is partially polarized perpendicularly to the plane of incidence, independently
to the polarization state of the incident photons. The degree
of polarization is \cite{Evans1977}:
\begin{equation}
{\cal P} = \frac{1-k}{1+k}. \label{eq:PolarizationDegree}
\end{equation}

\begin{figure}[htbp]
\begin{center}
\subfigure[\label{fig:k}]{\includegraphics[angle=0, totalheight=6cm]{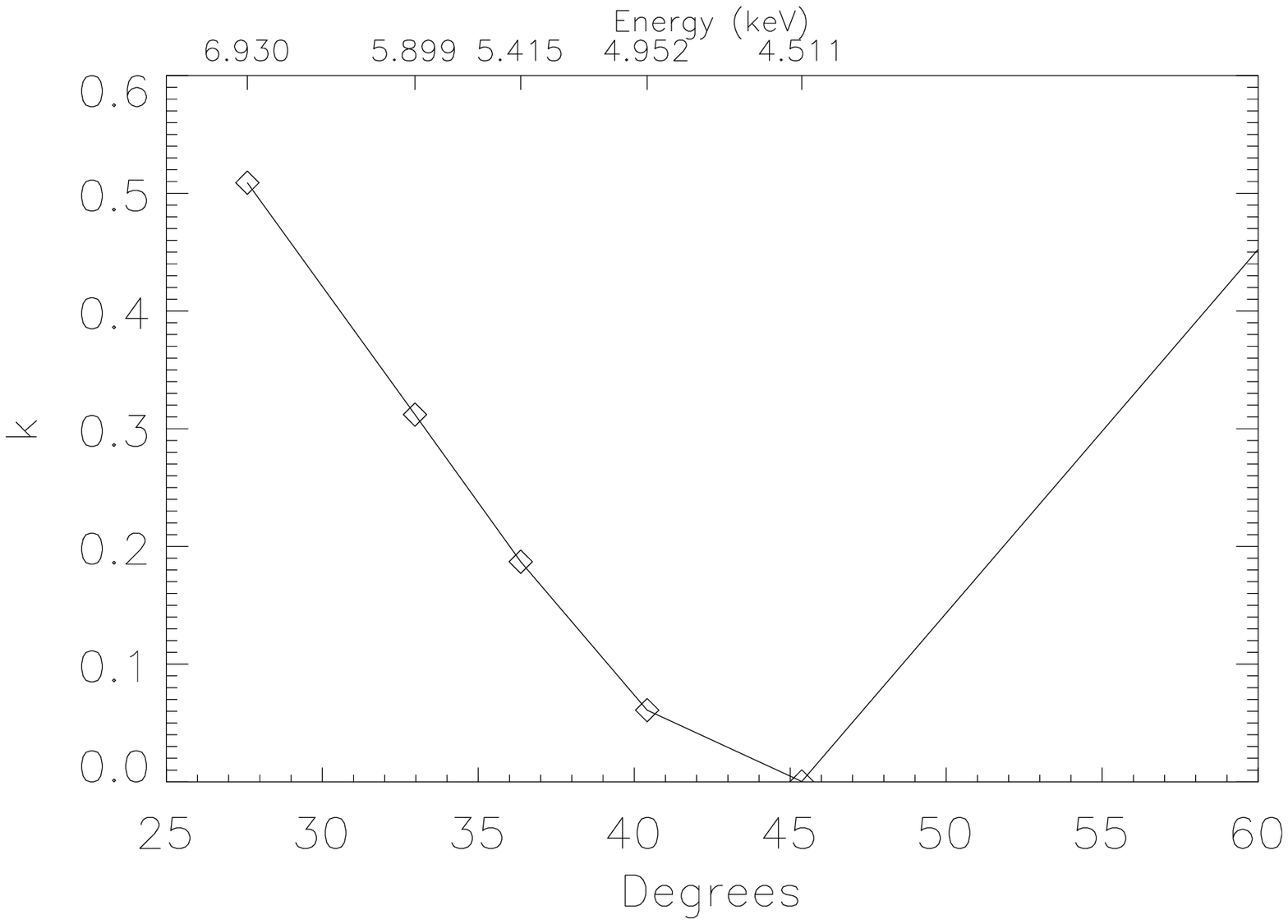}}
\subfigure[\label{fig:P}]{\includegraphics[angle=0, totalheight=6cm]{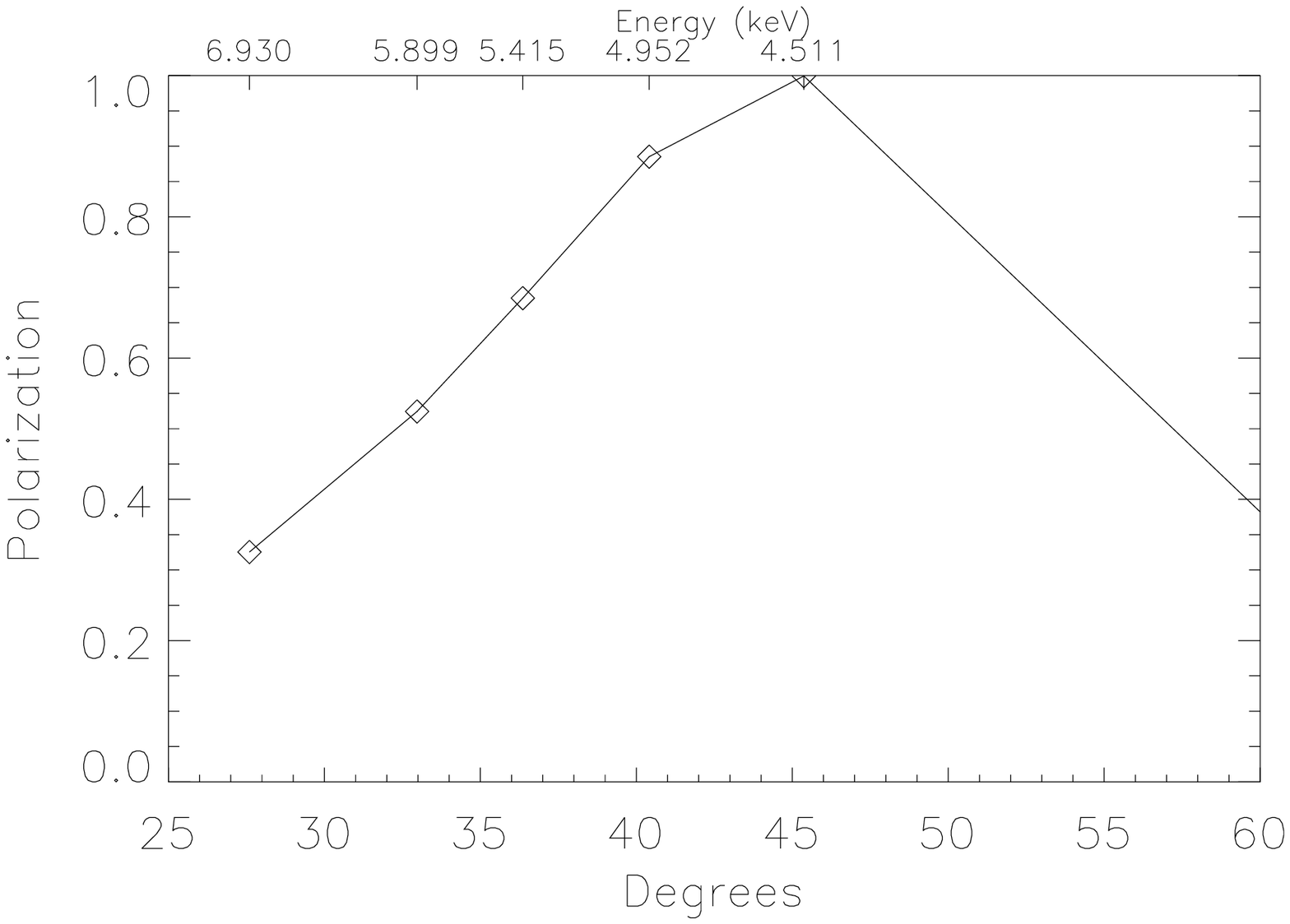}}
\end{center}
\caption{\small Polarization of the radiation diffracted on a fluorite crystal. ({\bf
a}) Dependence of $k$ with angle and then energy. ({\bf b}) Expected degree of polarization,
derived from Eq.~\ref{eq:PolarizationDegree}. The value of $k$ has been calculated by Henke et al.
\cite{Henke1993}.}
\end{figure}

In Fig.~\ref{fig:k} we report the ratio $k$ for the diffraction on the plane 002 of a fluorite
crystal (CaF$_2$), calculated by Henke et al.\cite{Henke1993}. For $\theta=45^\circ$, $k=0$ and,
according to Eq.~\ref{eq:PolarizationDegree}, the degree of polarization ${\cal P}=1$, as shown in
Fig.~\ref{fig:P}. Then the radiation is almost completely polarized when the diffraction angle is
constrained to nearly 45$^\circ$. Moreover, the diffracted photons must satisfy, within some eV for
flat crystals, the Bragg condition:
\begin{equation}
E = \frac{nhc}{2d\sin\theta}, \label{eq:BraggLaw}
\end{equation}
where $h$ and $c$ are respectively Planck's constant and the speed of light, $d$ the crystal lattice
spacing and $n$ the diffraction order. Hence, by constraining $\theta\approx45^\circ$, the
spectrum is composed of equally spaced, nearly monochromatic and polarized lines corresponding to
the various orders of diffraction $n$. Moreover, the choice of crystals with different lattice
spacing allows the production of lines at different energies.

If continuum radiation is diffracted, the quick change of $k$ with the diffraction angle can reduce 
the mean polarization of output radiation, since photons at slightly different energies are
diffracted at angles (and then with polarization) slightly different. Hence, very tight limits
around $\theta\approx45^\circ$ must be set with a collimator. However this limits the flux, since in
this case the diffracted photons must have an energy centered at the Bragg energy more or less some
tens of eV. A trade-off between the degree of polarization and reasonable fluxes can be then
achieved by choosing the degree of collimation.

Conversely, if line emission at energy close to the Bragg energy is employed, line photons are
diffracted with an efficiency of about 50\%, with a small contribution from the continuum photons
that can be generally neglected. Hence the precise diffraction angle, and consequently the degree of
polarization, can be easily derived from Eq.~\ref{eq:BraggLaw} and Eq.~\ref{eq:PolarizationDegree}
respectively, thanks to the knowledge of the energy of incoming line photons.

\subsection{The source} \label{sec:Source}

A source based on the Bragg diffraction was already presented by Muleri et al.\cite{Muleri2007}.
Here we present an improved version of that first prototype, with which we are able to produce
polarized photons starting at 1.6~keV with by far higher fluxes by exploiting medium-power X-ray
tubes and new crystals.

The improved version of the source shares the same base-design as the prototype. Lead-glass
capillary plates are used to collimate X-rays, since their 10~$\mu$m diameter holes, organized in a
hexagonal pattern with an on-axis transparency of 57\%, allow good collimation with limited size.
Semi-collimations of $\frac{1}{40}=1.4^\circ$ and $\frac{1}{100}=0.6^\circ$ are achieved by means of
capillary plates 0.4 and 1.0~mm thick, with an effective diameter of 20 and 27~mm respectively. The
type and the number of capillary plates (one or two, to constrain only the input or output
radiation or both) are chosen to achieve a trade-off between reasonable fluxes and high degree of
polarization.

Unpolarized X-rays are produced by commercial tubes with a maximum power of 50~W, manufactured by
Oxford Instruments. Three different tubes, with anode of molybdenum, rhodium and titanium, are
employed to produce line emission at 2.29, 2.69 and 4.51~keV respectively. These energies are in
accordance with the 45$^\circ$ Bragg diffraction from rhodium (001), germanium (111) and fluorite
(CaF$_2$, 220) crystals \cite{Henke1993}. Moreover, an ADP (101) and a PET (002) crystals are
employed to produce polarized emission at 1.65 and 2.04~keV by means of diffraction of continuum
radiation. In the table~\ref{tab:Crystals} the properties of the X-ray tubes and of the crystals
available in our facility are summarized.

\begin{table}[htbp]
\begin{center}
\begin{tabular}{l|c|c|c|c} 
Incident radiation (X-ray tube) & Energy (keV) & Crystal & $\theta$ & ${\mathcal P}$    \\
\hline
\hline
Continuum & 1.65 & ADP (NH$_4$H$_2$PO$_4$, 101)& 45$^\circ$ & $\sim$1.0 \\
Continuum & 2.04 & PET (C$($CH$_2$OH$)_4$, 002)& 45$^\circ$ & $\sim$1.0 \\
L$\alpha$ Molybdenum (50~W)& 2.293 & Rhodium (001) & 45.36$^\circ$ & 0.9994 \\
Continuum & 2.61 & Graphite (002) & 45$^\circ$ & $\sim$1.0 \\
L$\alpha$ Rhodium (50~W) & 2.691 & Germanium (111) & 44.86$^\circ$ & 0.9926 \\
K$\alpha$ Calcium (200~mW) & 3.692 & Aluminum (111) & 45.88$^\circ$ & 0.9938 \\
K$\alpha$ Titanium (50~W)& 4.511 & Fluorite CaF$_2$ (220) & 45.37$^\circ$ & 0.9994 \\
\end{tabular}
\caption{X-ray tubes and crystals available in our facility. The low power calcium tube is
employed with the first prototype of source but it is shown in Table because can be mounted on
the mechanical assembly described in Sec.~\ref{sec:MechanicalAssembly}. For the
diffraction on ADP, PET and graphite crystals, continuum radiation produced by any tube can be
employed since no line emission in accordance with the diffraction at 45 degrees is
available. In this case, the energy and the polarization in Table are referred to the photons
diffracted at 45$^\circ$. Only the energies corresponding to the first order of diffraction are
reported, but each crystal can produce polarized radiation even at higher orders and, hence, at
energies which are integer multiple of the first order. Polarization from calculation
performed by Henke et al.\cite{Henke1993}.} \label{tab:Crystals}
\end{center}
\end{table}

\begin{figure}[htbp]
\begin{center}
\subfigure[\label{fig:Stage}]{\includegraphics[angle=0, totalheight=6cm]{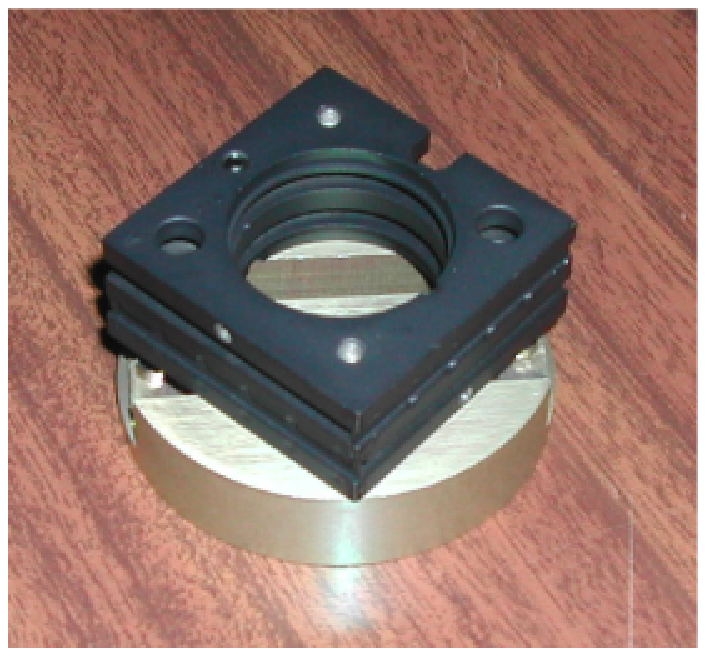}}
\subfigure[\label{fig:Polarizer}]{\includegraphics[angle=0, totalheight=6cm]{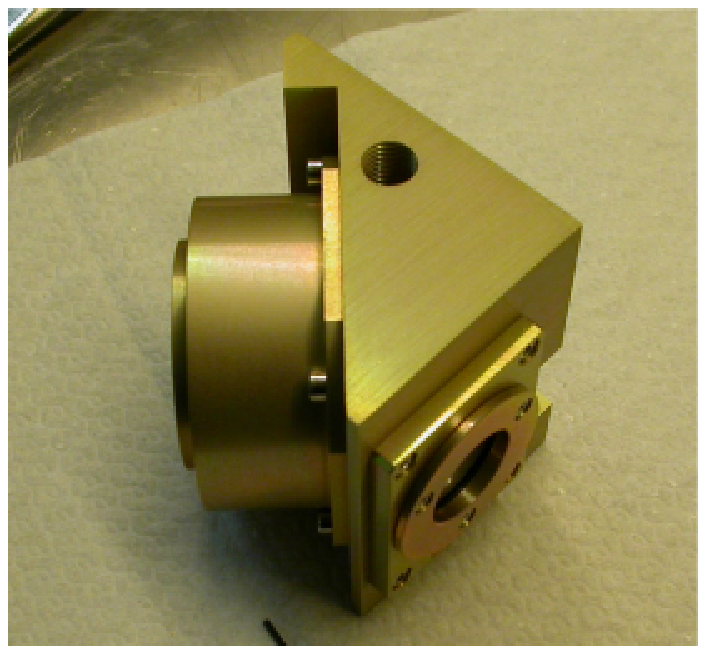}}
\end{center}
\caption{\small ({\bf a}) The manual stage on which each crystal is mounted. It allows two axes
tilt regulation in the range $\pm3^\circ$ and then an optimum alignment to the Bragg condition. The
aluminum case of the crystal is below the stage. ({\bf b}) The crystal mounted on the central body
of the source. The stage which allows the tilt regulation is hidden in the cover on the left. A
capillary plate, used as collimator, is visible on the right in its case. The threaded hole on the
top is for helium flowing to reduce the air absorption.}
\end{figure}

Since the choice of the diffracting crystal allows to produce polarized radiation at different
energies, we mounted each crystal in a different aluminum case to easily manage and change them. An
important difference between the prototype and the improved version of the source is that in the
latter we mounted each crystal case on a manual stage which allows two axes tilt regulation in the
range $\pm3^\circ$ (see Fig.~\ref{fig:Stage}). The stage is then mounted on a central body (see
Fig.~\ref{fig:Polarizer}) and it is regulated to achieve the best alignment to the Bragg condition
by means of optical and X-ray measurements. In the first case, a mirror mounted on the case of the
crystal is employed to reach the optimum angle between the crystal and the collimators, which are
partially reflective. Then the inclination is refined with X-ray measurement to reach the highest
flux.

\begin{figure}[htbp]
\begin{center}
\subfigure[]{\includegraphics[angle=0, totalheight=7cm]{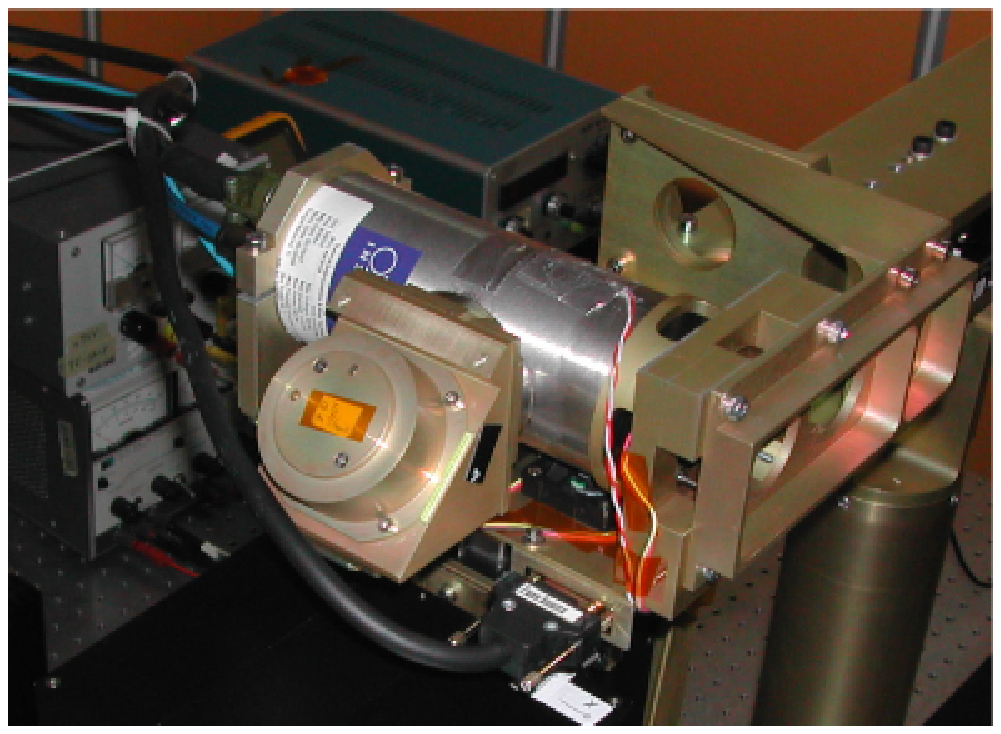}}
\subfigure[]{\includegraphics[angle=0, totalheight=7cm]{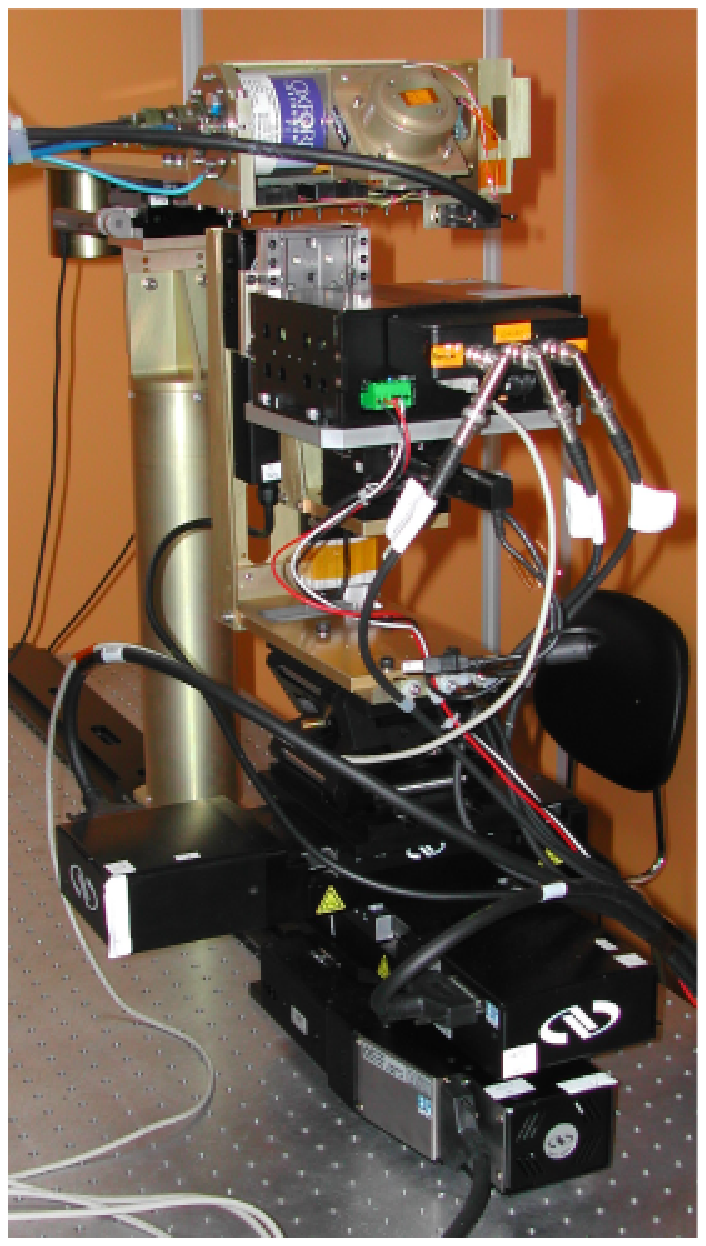}}
\end{center}
\caption{\small The polarized source mounted on the mechanical assembly described in
Sec.~\ref{sec:MechanicalAssembly}. \label{fig:Source}}
\end{figure}

We added two standard gas connectors in the central body (see Fig.~\ref{fig:Source}) on the top and
bottom side of the source for helium flowing to reduce air absorption which heavily affects low
energy photons. For the same reason, we reduced to the minimum the path of the photons.

The collimators also are mounted on an aluminum case, since their choice allows to achieve a
trade-off between high fluxes and clean spectra. Two issues can indeed detach the output radiation
from the ideal condition of monochromatic and highly polarized photons. The first one is that
radiation can be scattered at nearly 90$^\circ$ on the crystal or on its case and then photons not
at the Bragg energy can emerge from the source. The second issue is that the input radiation is
partially absorbed in the crystal (and in the aluminum case) and then unpolarized
fluorescence emission is produced. This problem is especially important when 1.65~keV
polarized photons are produced, since the aluminum K$\alpha$ fluorescence is at 1.5~keV, and with
the fluorite crystal, since the 3.7 K$\alpha$ fluorescence of calcium has a comparable
energy (and hence absorption) with respect to the polarized photons at 4.5~keV produced with
diffraction on that crystal.

\begin{figure}[htbp]
\begin{center}
\subfigure[]{\includegraphics[angle=90, width=7cm]{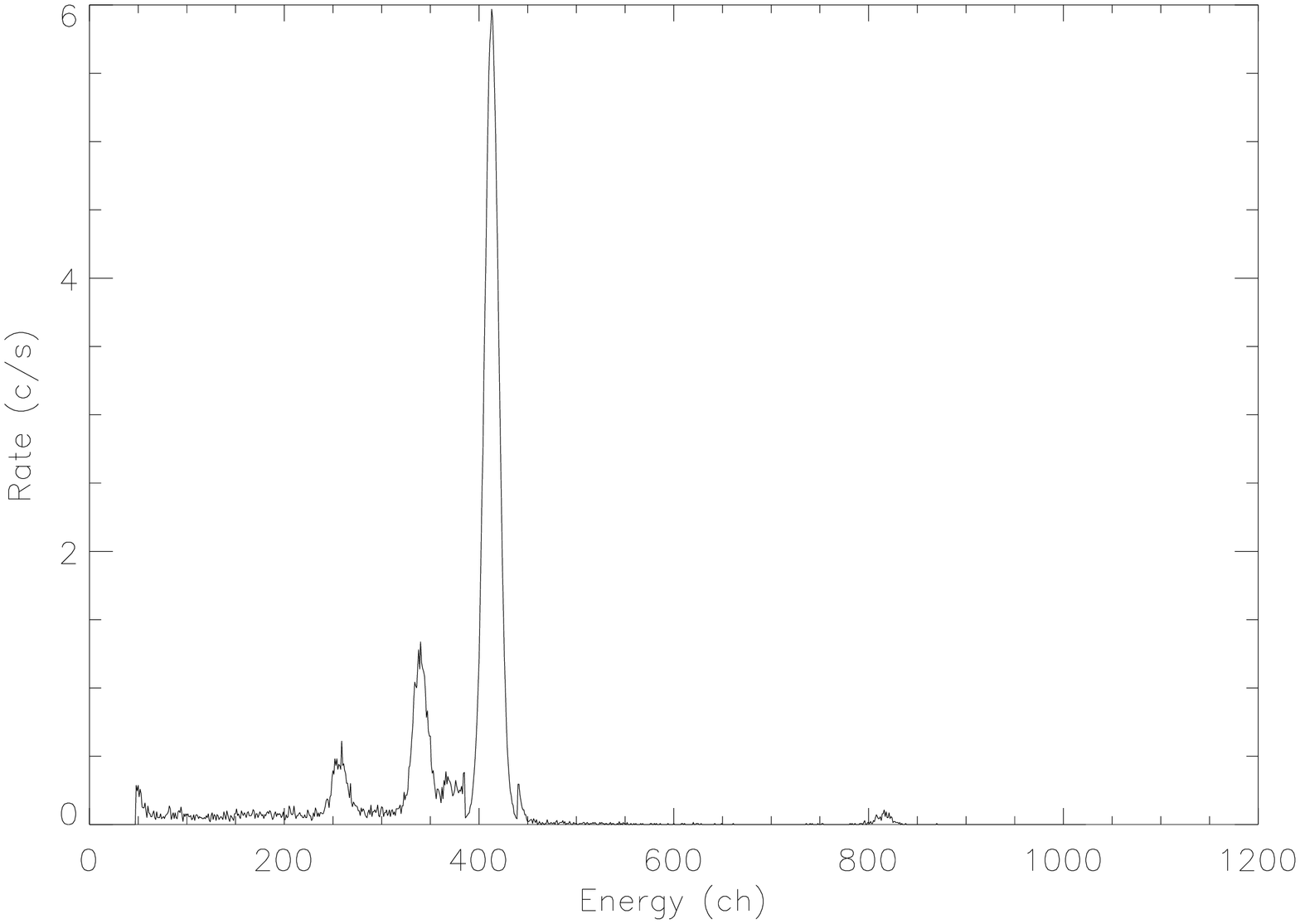}}
\subfigure[]{\includegraphics[angle=90, width=7cm]{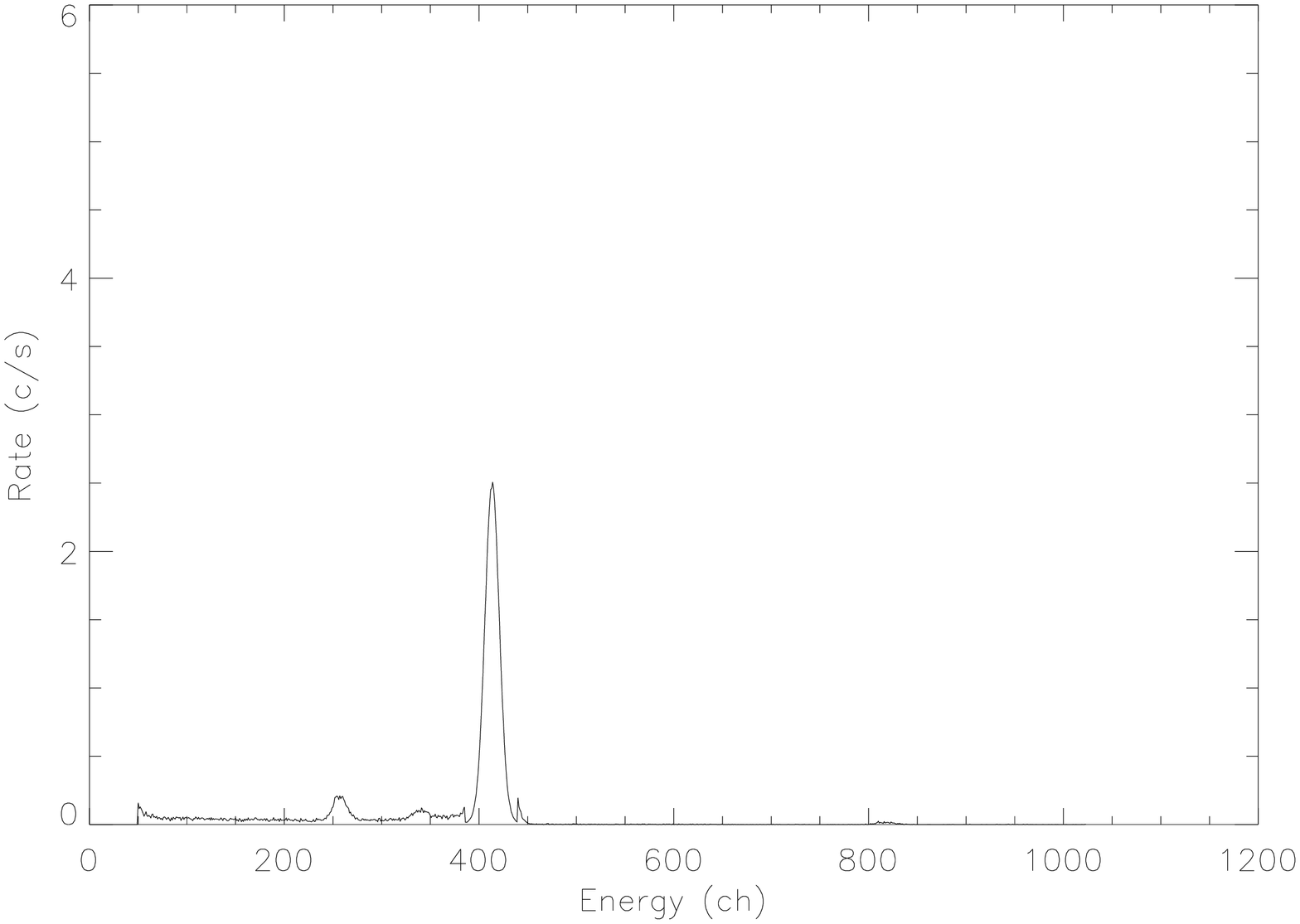}}
\end{center}
\caption{\small Spectrum of the polarized source when the X-ray tube with titanium anode and the
fluorite crystal are employed. The spectrum is measured with an Amptek XR100CR Si-PIN detector, with
213~eV resolution at 5.9~keV. Note that the flux of the line at 4.5~keV (at about channel 410) has
been multiplied by 0.1. The peak in the middle is the fluorescence of the calcium, while the peak on
the left, at about 250~ch, is the escape peak of the line at 4.5~keV from the silicon detector.
({\bf a}) One collimator with $\frac{1}{40}$ is employed to constrain the direction of diffracted
photons. ({\bf b}) The same as ({\bf a}), but when a $\frac{1}{40}$ and a $\frac{1}{100}$
collimators are employed to constrain the direction of input and output radiation respectively
\label{fig:Spectra}.}
\end{figure}

The flux of fluorescence and of the scattered radiation can be reduced with respect to the polarized
one by means of narrower collimators, which reduce the incident flux but save the Bragg condition
for diffraction. Indeed the first two components are approximately proportional to the incident
radiation, while the latter depends only on the flux of the line emission. For example, in
Fig.~\ref{fig:Spectra} we report the spectra from the diffraction of emission of the tube with
titanium anode on the fluorite crystal with different configurations of collimators, while the
ratio between the fluxes of the polarized line and that of the fluorescence of calcium at 3.7~keV is
reported in Table~\ref{tab:Fluxes}.

\begin{table}[htbp]
\begin{center}
\begin{tabular}{l|c|c|c|c} 
Input collimator & Output collimator & Flux$_{polarized}$ (c/s) & Flux$_{fluorescence}$ (c/s) &
Ratio \\
\hline
\hline
--- & $\frac{1}{40}$ & 1073.7$\pm$3.0 & 18.04$\pm$0.47 & $\sim$60 \\
$\frac{1}{40}$ & $\frac{1}{100}$ & 458.38$\pm$0.79 & 1.046$\pm$0.072 & $\sim$440 \\
\end{tabular}
\caption{Ratio between the fluxes of the polarized line at 4.5~keV and that of the fluorescence one
at 3.7~keV for the same crystal and X-ray tube, but for different configurations of collimators.}
\label{tab:Fluxes}
\end{center}
\end{table}

The complete source, composed by the polarizer and the X-ray tube, is shown in Fig.~\ref{fig:Source}
mounted on the mechanical assembly described in Sec.~\ref{sec:MechanicalAssembly}. The cooling of
the tube is performed with four fans and, if its temperature increases above a determined threshold,
it is turned off.

We have focused our attention to the first order of diffraction, since in this case the line
emission of the tubes is in accordance with the diffraction at nearly 45$^\circ$. However the
continuum radiation produced by the X-ray tube by bremsstrahlung can be employed to produce also
high energy and polarized radiation exploiting higher orders of diffraction, i.e. at energies
which are integer multiple of the energy of the first order (see Table~\ref{tab:Crystals}). The
medium power X-ray tubes can be very efficiently employed at this aim, since the high voltage (and
hence the maximum energy of the photons produced) can be controlled independently from the current
(and hence from the flux of the photons). Then, to produce polarized photons at 9~keV exploiting
the second order of diffraction from fluorite, we can set the high voltage at 13~kV, i.e. just
below the energy of the third order of diffraction, and then change the current to increase the
flux up to 1~mA. 

\begin{figure}[htbp]
\begin{center}
\includegraphics[angle=90, totalheight=7cm]{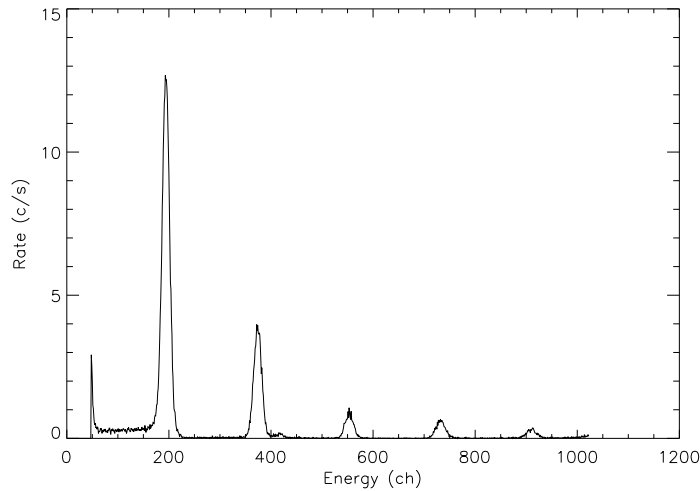}
\end{center}
\caption{\small Continuum radiation diffracted on the PET crystal. Input radiation is obtained with
the X-ray tube with titanium anode (15~kV, 0.5~mA). The first five orders of diffraction are
visible. The low energy peak (at about channel 50) is due to background in the detector.
\label{fig:HigherOrders}}
\end{figure}

In Fig.~\ref{fig:HigherOrders} we show the spectrum obtained with diffraction of continuum
radiation on the PET crystal. The first five orders are visible with a relative peak
heights which depend mainly on the different efficiency of the diffraction for each order and on
the spectrum of the unpolarized incident radiation. Helium flowing was employed to avoid absorption
of low energy photons. The rate of the lines are reported in Table~\ref{tab:PET_Fluxes}. 

\begin{table}[htbp]
\begin{center}
\begin{tabular}{l|l||l|l} 
  & Flux (c/s) &  & Flux (c/s) \\
\hline
\hline
2.04~keV - I order & 224.0$\pm$1.5 & 8.16~keV - IV order & 13.44$\pm$0.43 \\
4.08~keV - II order & 73.5$\pm$1.0 & 10.20~keV - V order & 7.41$\pm$0.27 \\
6.12~keV - III order & 17.10$\pm$0.39 \\
\end{tabular}
\caption{Main characteristics of the spectrum obtained with the polarized source by diffracting
continuum radiation on the PET crystal.} \label{tab:PET_Fluxes}
\end{center}
\end{table}

%

\section{The mechanical assembly} \label{sec:MechanicalAssembly}

The polarized source described in Sec.~\ref{sec:Source} allows to overcome the problem of the
measurement of the low energy performances of the GPD, since photons with energy as low as 1.65~keV
can be produced. The next improvement is then a systematic study of the sensibility of the GPD with
a space-controlled beam. In particular, we want to explore:
\begin{itemize}
\item the response on all the surface of the detector, since the GPD can also perform imaging on
extended sources (XY-mapping);
\item the correlation between the direction and the reconstructed angle of polarization (angle
reconstruction);
\item the study of the response to inclined beams, due to the focusing of grazing incidence optics
or to large field of view instruments (inclined measurements).
\end{itemize}

We employed the Bragg diffraction sources described above and high precision motorized stages,
manufactured by Newport, to built a facility to perform these tasks. In addition to the Bragg
sources
(the prototype and the improved version), we interfaced even unpolarized sources with the facility.
In summary, we can produce:
\begin{itemize}
\item a collimated and, in case, diaphragmed \emph{polarized} and nearly monochromatic beam,
produced by means of Bragg diffraction. We can employ the prototype source at 3.7~keV or the
improved version at 1.65, 2.04, 2.29, 2.69 and 4.51~keV. The higher orders of diffraction are also
available.
\item collimated and/or diaphragmed \emph{unpolarized} radiation at 5.89 and 6.49~keV by means of
a Fe$^{55}$ radioactive source.
\item collimated and/or diaphragmed \emph{unpolarized} radiation at about 2.29, 2.69, 4.51~keV by
means of direct emission from X-ray tubes.
\end{itemize}

\begin{figure}[ht]
\begin{center}
\includegraphics[angle=0,totalheight=10cm]{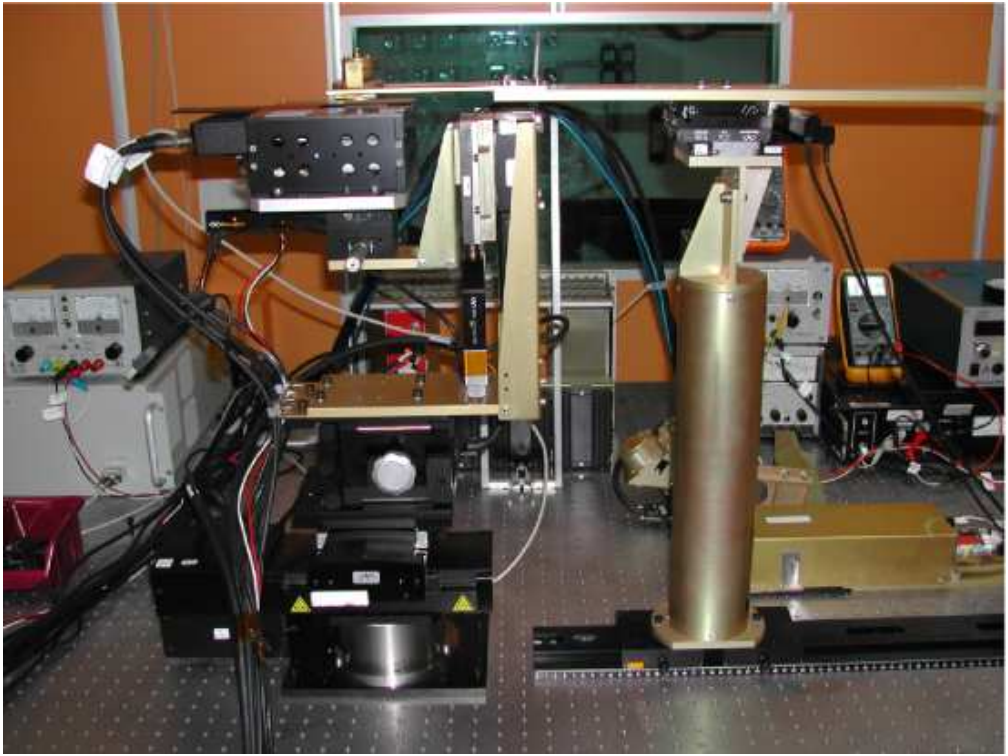}
\end{center}
\caption{A picture of the complete facility, with the source on the right and the tower
supporting the detector on the left. As shown, the source is mounted on a rail. In this case,
the Fe$^{55}$ radioactive source and the GPD are mounted. \label{fig:BabylonTowerComplete}}
\end{figure}

The mechanical assembly is designed to perform automatically and subsequently the measurements
listed above, i.e. XY-mapping, angle reconstruction and inclined measurements, for each source. The
detector is mounted on a platform which can be rotated, inclined and moved with respect to the beam
with eight motorized stages and two manual ones. Moreover, the source is mounted on a rail with a
graduated scale which allows its quick movement when maintenance operations are performed on the
tower supporting the detector (see fig.~\ref{fig:BabylonTowerComplete}). The facility is mounted on
an optical table in a room, shielded by X-ray radiation with a 1~mm of lead.

\begin{figure}[ht]
\begin{center}
\includegraphics[angle=0, totalheight=12cm]{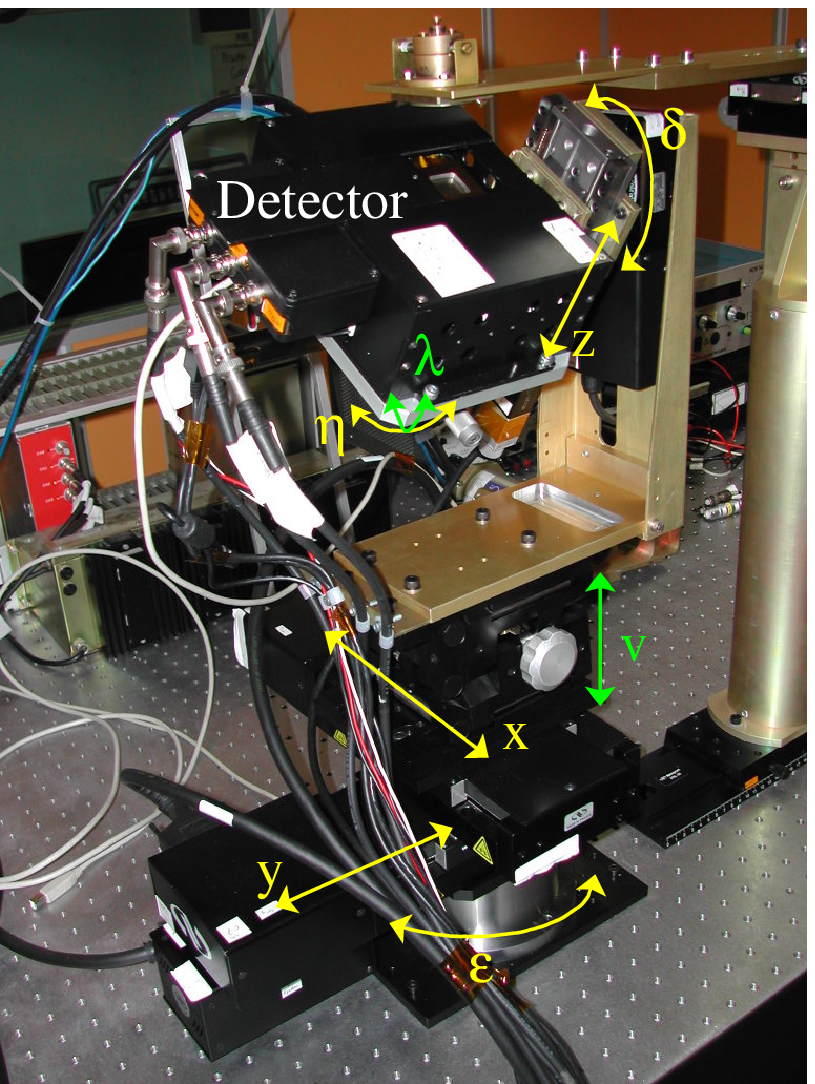}
\end{center}
\caption{The name and the role of each stage. In yellow are the motorized stages, while in green are
the manual ones. \label{fig:BabylonTower}}
\end{figure}

The stages are divided into two groups. The first one includes five motorized stages which actually
change the direction of the detector with respect to the beam, while the second group is employed
to align the beam with the detector. In the former group there are (see
fig.~\ref{fig:BabylonTower}):

\begin{enumerate}
\item$\delta$ (\emph{inclination angle}, resolution 1''): The main purpose of this stage is to allow
the inclination
of the detector at large angles, typically between +60 and -60 degrees. Moreover it is employed with
$\eta$ to automatically adjust the tilt angle of the detector, i.e. to align the detector
orthogonally to the incoming beam.
\item$\epsilon$ (\emph{polarization angle}, resolution 0.2''): This stage allows to change the
direction of
polarization of the beam. The range is between 0 and 360$^\circ$. A dedicated control has been
implemented in the software to prevent the twisting of the cables.
\item$x$: (resolution 0.5~$\mu$m) This stage allows the XY mapping together with $y$. Moreover it is
employed to center the
detector with the beam.
\item$y$: (resolution 0.5~$\mu$m) The role of this stage is complementary to $x$.
\item$z$: This stage is employed to adjust the distance between the
detector and the source. While,
at low energy, the beam is heavily affected by air absorption and so the distance from the source
must be reduced, during the measurements at large angles of inclination the detector should be
placed on the rotation axis of $\delta$, to prevent the movement of the detector with the increasing
of the inclination angle. Hence this stage for all practical purposes has two positions, \textsc{Up}
when the distance from the source is reduced, and \textsc{Down} when the axis of rotation of the
inclination crosses the detector.
\end{enumerate}

The stages which instead are employed to align the detector with the beam are:
\begin{enumerate}
\setcounter{enumi}{5}
\item$\eta$: This motorized stage allows to tilt the plane of the detector to align it
orthogonally to the incoming beam.
\item$xso$ (resolution 0.1~$\mu$m): The purpose of this motorized stage is to center, together with
$yso$, the beam with the
axis of rotation of the stage $\epsilon$. This allows to study how the angle of polarization is
measured in each region of the detector and if systematic effects exist.
\item$yso$ (resolution 0.1~$\mu$m): The role of this motorized stage is complementary to $xso$.
\item$\lambda$: This stage allows a manual tilt regulation of the plane of the detector. For all
practical purposes this stage is redundant with respect to $\delta$.
\item$\chi$: The rail which allows the quick movement of the source when maintenance operation
are performed on the tower. The rail is provided with a graduated scale with 1~mm pitch.
\item[---.] $\nu$: A further stage allows a vertical adjustment with a range of about 44~mm.
\end{enumerate}

Hence, a total of ten stages are employed to align and control the beam. All the motorized ones
are connected to a controller, which in its turn is connected with a PC by means of an ethernet
cable. The motorized stages can be manually moved with a remote control or by means of a web
interface of the controller.

\begin{figure}[ht]
\begin{center}
\subfigure[\label{fig:LabView1}]{\includegraphics[angle=0, totalheight=7cm]{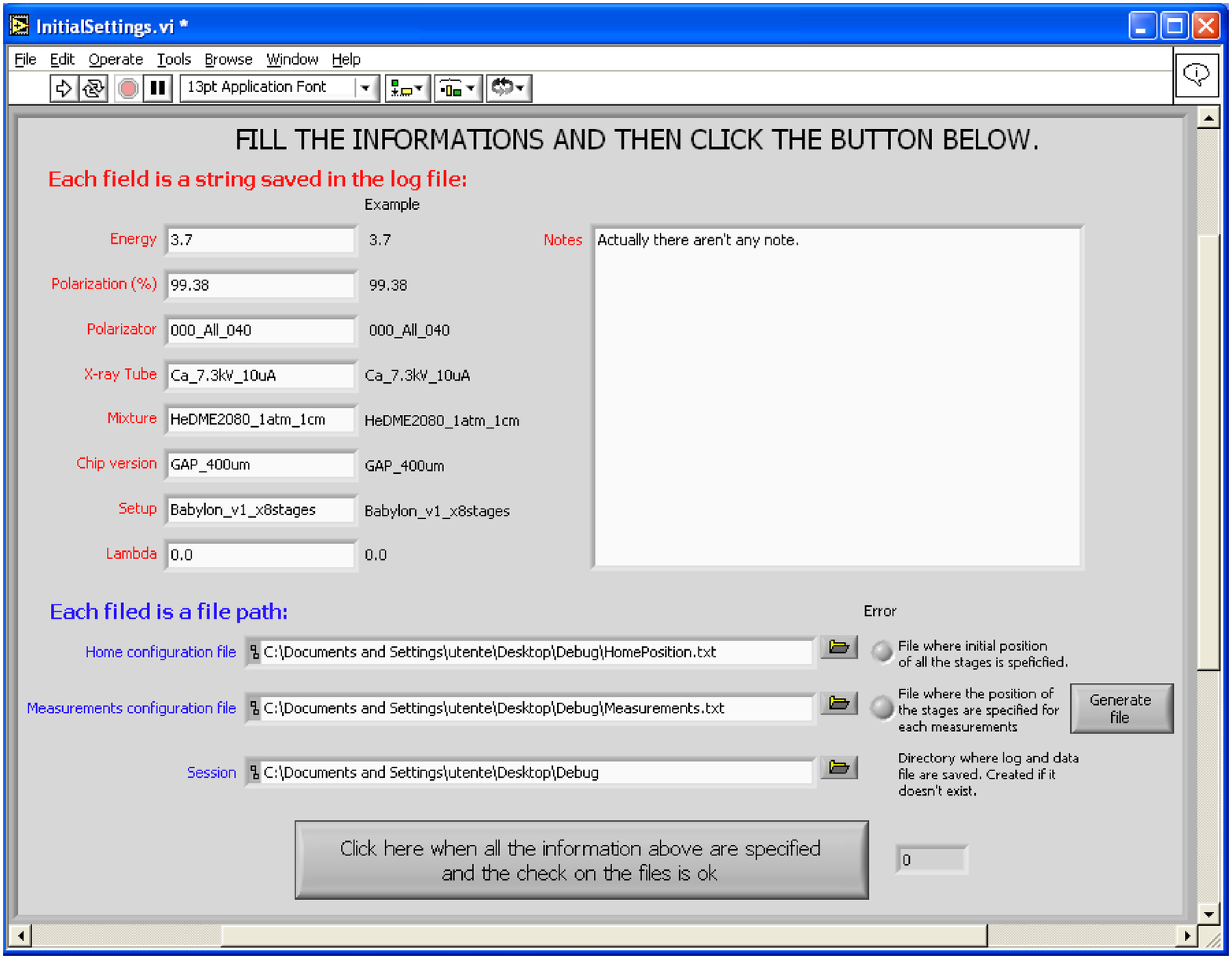}}
\subfigure[\label{fig:LabView2}]{\includegraphics[angle=0, totalheight=7cm]{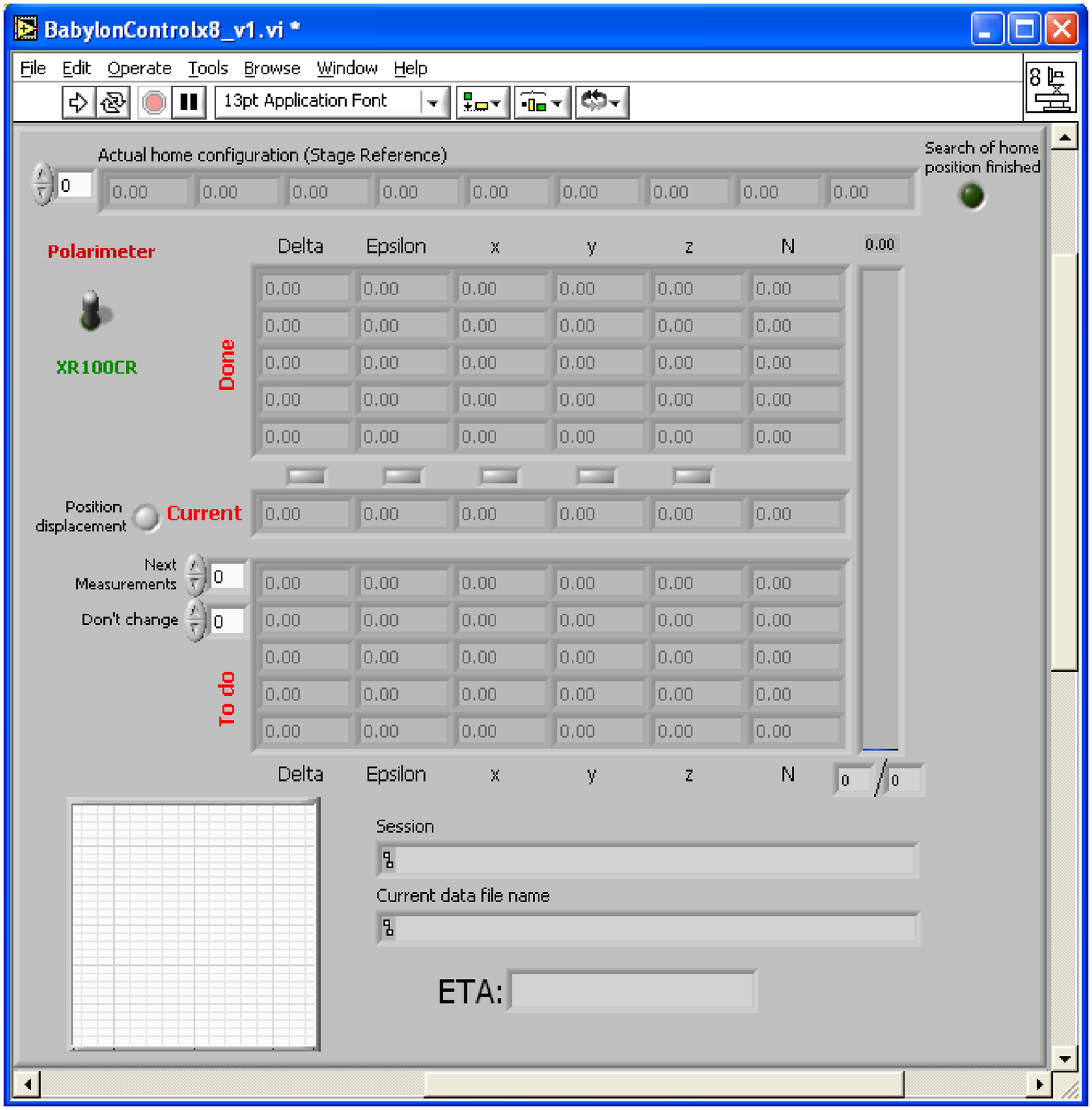}}
\end{center}
\caption{User interface of the program which controls the motorized stages. ({\bf a}) Interface
in which all the informations are at first collected. ({\bf b}) The panel which allows to control
the state of the measurements.}
\end{figure}

The movements of the facility and the acquisition of the GPD are controlled by means of a LabView
software, which can manage all kind of studies quoted above to perform long-lasting sessions of
sequentially measurements, without the presence of the user. The control program is
organized into two steps. All the informations required to the session of measurements are
collected in the first one (see Fig.~\ref{fig:LabView1}), after that a second panel reports
the current measurement and the progress of the session (see fig.~\ref{fig:LabView2}). The positions
of all stages are saved in a log and in the data file for subsequent automatic analyzes.

\section{Preliminary results}

A test of the facility was performed by using an Amptek XR100CR Si-PIN detector to map the 3.7~keV
K$\alpha$ calcium line diffracted by the aluminum crystal. A diaphragm of 1~mm is used in front of
the detector to sample the beam.

\begin{figure}[ht]
\begin{center}
\includegraphics[angle=90, width=11cm]{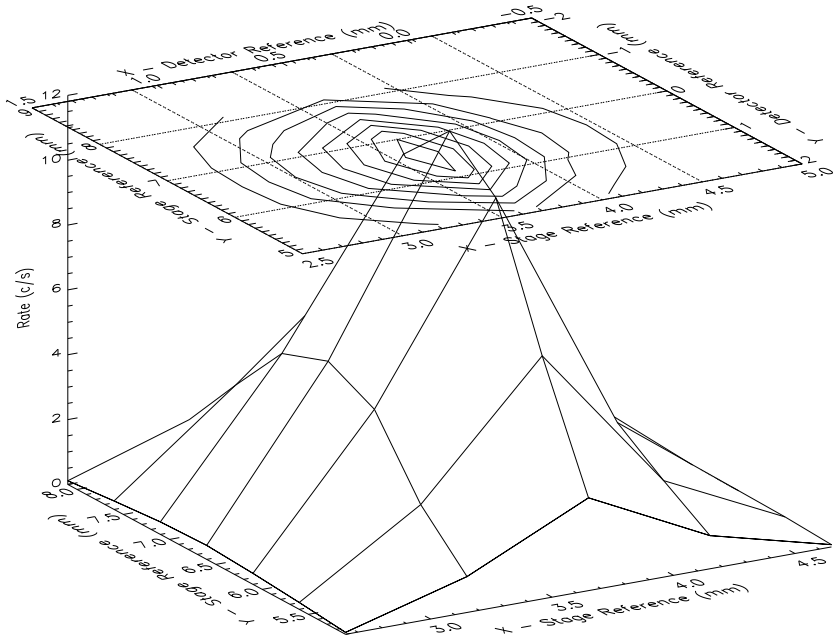}
\end{center}
\caption{Map (7$\times$5 points) of the the rate of the beam produced with the Bragg diffraction at
3.7~keV. The beam is elliptical, with a major axis (in the direction of polarization) of
about 4~mm FWHM.}
\label{fig:Plot_Rate}
\end{figure}

In Fig.~\ref{fig:Plot_Rate} the map of the rate of the flux is shown on a grid of 7$\times$5
points. The beam is elliptical, with a major axis (in the direction of polarization) of about 4~mm
FWHM. In the perpendicular direction, the extension of the beam is 2~mm FWHM. Since the diffraction
should not change the section of the beam, it should reflect the shape of cathode of the X-ray tube.

\begin{figure}[ht]
\begin{center}
\subfigure[\label{fig:Plot_FWHM}]{\includegraphics[angle=90,
width=7cm]{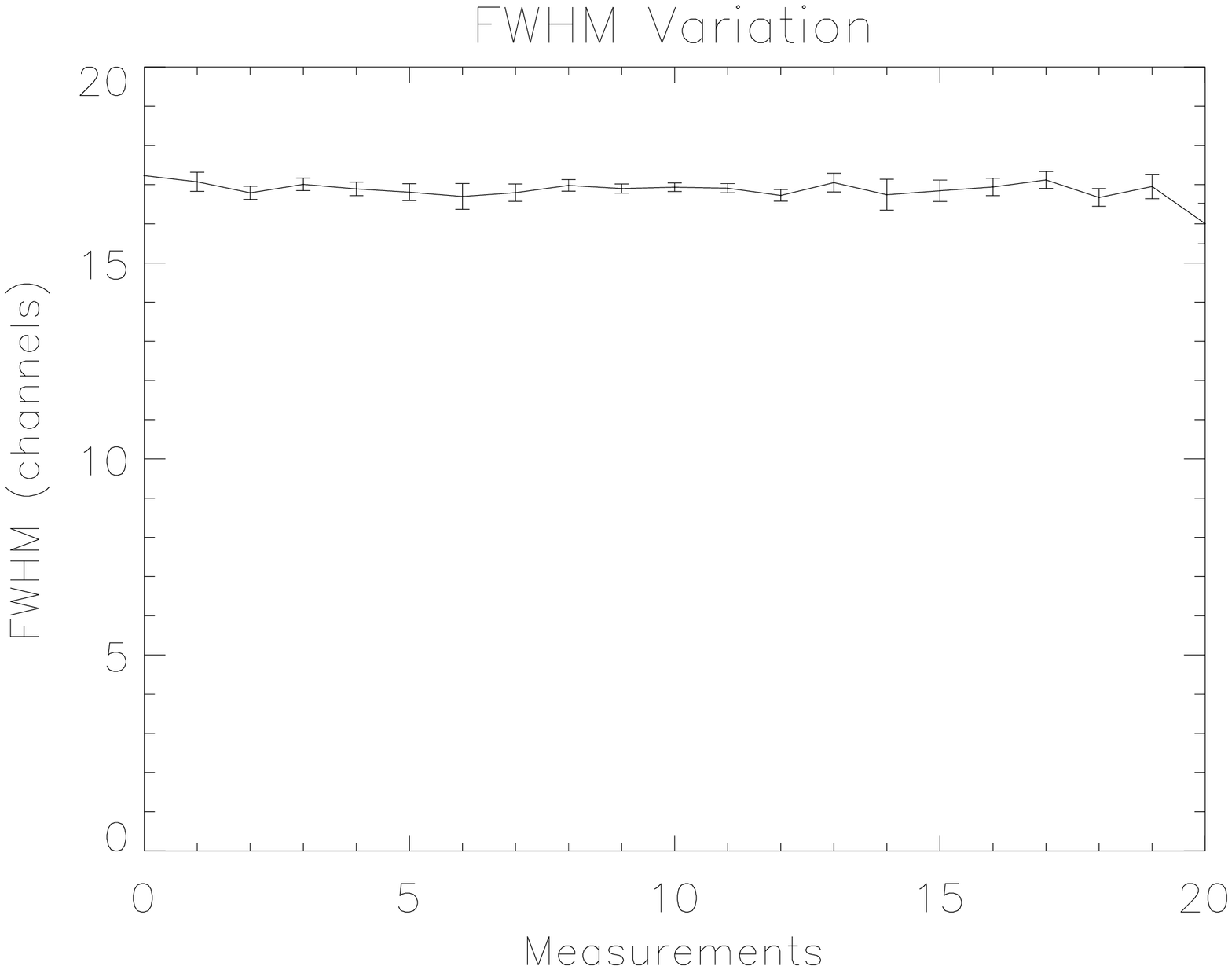}}
\subfigure[\label{fig:Plot_LineEnergy}]{\includegraphics[angle=90,
width=7cm]{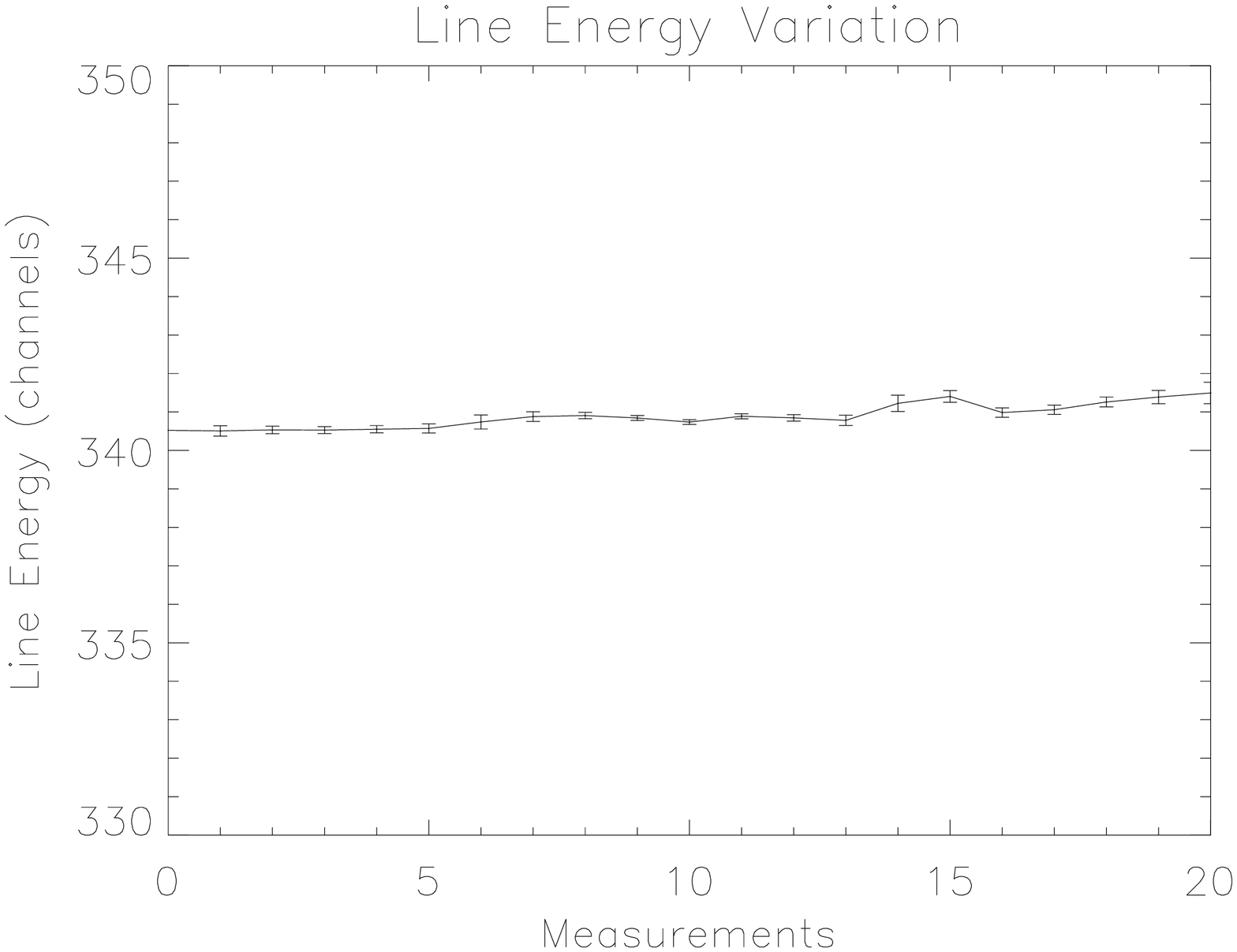}}
\end{center}
\caption{FWHM ({\bf a}) and energy ({\bf b}) variation of the 3.7~keV K$\alpha$ calcium line
diffracted by the aluminum crystal and measured with an Amptek XR100CR Si-PIN detector. In the
x-axis, the index of the 7$\times$3=21 measurements is reported.}
\end{figure}

The properties of the line are identical for all the central 7$\times$3 points, where the counts
are sufficient for a reliable fit. In Fig.~\ref{fig:Plot_FWHM} we report the FWHM of the line
for the 7$\times$3 measurements, which are clearly consistent within errors. In the
Fig.~\ref{fig:Plot_LineEnergy} instead, the energy of the line is shown. In this case a increasing
trend is present, even if it can be imputed to changes in the silicon detector, since it follows the
sequence in which the measurements were done.

\section{Conclusion}

We have designed and built a facility for the calibration of the next generation X-ray
polarimeters. Polarized photons are produced by means of Bragg diffraction at nearly 45 degrees.
Following the good results of a prototype source which allows the diffraction of
continuum or line emission on aluminum or graphite crystals\cite{Muleri2007}, we have developed an
improved version of the source which can exploit also the diffraction on ADP, PET, rhodium,
germanium and fluorite crystals. Thanks to X-ray tubes with anodes tuned with these crystals, we can
very efficiently produce almost completely polarized photons at 2.29, 2.69 and 4.51~keV. Moreover,
continuum emission is employed to generate highly polarized radiation at 1.65 and 2.04~keV.
Both the prototype and the improved version of the polarized sources have been interfaced,
together with unpolarized sources, with a mechanical assembly which, thanks to motorized stages
controlled via PC, allows calibration measurements with space-controlled beams. In particular the
facility is optimized for mapping the response of imaging devices to both polarized and unpolarized
radiation, with inclined or orthogonal beams. This facility will be soon employed for the
calibration of the Gas Pixel Detector.

\section*{Acknowledgments}

FM acknowledges financial support from Agenzia Spaziale Italiana (ASI)
under contract ASI~I/088/06/0.

\bibliography{References}   
\bibliographystyle{spiebib}   

\end{document}